\documentclass[%
 aip,
 amsmath,amssymb,
preprint,%
]{revtex4-1}
\usepackage{graphicx}% Include figure files
\usepackage{dcolumn}% Align table columns on decimal point
\usepackage{bm}% bold math
\usepackage{xcolor}
\usepackage[utf8]{inputenc}
\usepackage[T1]{fontenc}
\usepackage{mathptmx}
\usepackage{mathtools}
\usepackage[font=small,labelfont=bf,
   justification=raggedright,
   format=plain]{caption}
\usepackage{subcaption}
\usepackage{setspace}
\usepackage{amsmath}

\begin{document}
\preprint{AIP/123-QED}
\setstretch{1.9} 

% TITLE AND AUTHORS
\title{Slip and Stress From Low Strain-Rate Nonequilibrium Molecular Dynamics: The Transient-Time Correlation Function Technique}
\author{Luca Maffioli}
\affiliation{Department of Mathematics, School of Science, Computing and Engineering Technologies, Swinburne University of Technology, PO Box 218, Hawthorn, Victoria 3122, Australia}
\author{Edward R. Smith}
\affiliation{Mechanical and Aerospace Engineering, Brunel University London, Kingston Lane, Uxbridge UB8 3PH, United Kingdom}
\author{James P. Ewen}
\affiliation{Department of Mechanical Engineering, Imperial College London, London SW7 2AZ, United Kingdom}
\author{Peter J. Daivis}
\affiliation{School of Science, RMIT University, GPO Bpx 2476, Victoria 3001, Australia}
\author{Daniele Dini}
\affiliation{Department of Mechanical Engineering, Imperial College London, London SW7 2AZ, United Kingdom}
\author{B. D. Todd}
\email{btodd@swin.edu.au}
\affiliation{Department of Mathematics, School of Science, Computing and Engineering Technologies, Swinburne University of Technology, PO Box 218, Hawthorn, Victoria 3122, Australia}

\date{\today}% It is always \today, today,
%\begin{document}
%\maketitle

% ABSTRACT
\begin{abstract}
We derive the transient-time correlation function (TTCF) expression for the computation of phase variables of inhomogenous confined atomistic fluids undergoing boundary-driven planar shear (Couette) flow at constant pressure. Using nonequilibrium molecular dynamics simulations, we then apply the TTCF formalism to the computation of the shear stress and the slip velocity for atomistic fluids at realistic low shear rates, in systems under constant pressure and constant volume. We show that, compared to direct averaging of multiple trajectories, the TTCF method dramatically improves the accuracy of the results at low shear rates, and that it is suitable to investigate the tribology and rheology of atomistically detailed confined fluids at realistic flow rates.  
\end{abstract}
\maketitle

% INTRODUCTION
\section{Introduction}
Nonequilibrium molecular dynamics (NEMD) simulations have given unique insights into nanoscale fluid behaviour in a range of applications from tribology\cite{Ewen2018} to hydraulic fracturing and polymer processing \cite{doi:10.1021/acs.jpclett.6b01684, Gartner2019}. On the other hand, NEMD simulations are generally limited to external fields which are several orders of magnitude larger than those encountered in experiments and applications. For example, typical shear rates\cite{doi:10.1177/1350650117696181} in automotive engines range from $10^5-10^8\;\text{s}^{-1}$, whilst those in polymer processing and hydraulic fracturing are usually between $10^3-10^6\;\text{s}^{-1}$ and $1-10^{3}\;\text{s}^{-1}$, respectively \cite{Vlachopoulos2003, doi:10.1146/annurev-chembioeng-080615-033630}. Even using massively-parallelized MD software\cite{PLIMPTON19951, THOMPSON2022108171} on modern high performance computer (HPC) systems, the lowest accessible shear rates by direct NEMD\cite{Jadhao7952} are $>10^5\;\text{s}^{-1}$.

For atomic fluids, the viscosity remains Newtonian up to high shear rates\cite{9943764}; however, for molecular fluids, non-Newtonian shear thinning behaviour can dominate the viscous response at shear rates well below those accessible to direct NEMD simulations\cite{Jadhao7952,PhysRevLett.88.058302}. Comparison of NEMD results with experimental viscosity measurements is thus generally restricted to extrapolation\cite{Jadhao7952,C7CP01895A} or time-temperature superposition\cite{Jadhao7952,PhysRevLett.88.058302} methods.

One approach which allows the study of fluid rheology at realisitic shear rates is the transient-time correlation function (TTCF) technique \cite{PhysRevA.10.2461,Dufty1979,COHEN198317,PhysRevA.35.792,PhysRevA.38.4142}, which is based on the time-correlation between the initial rate of energy dissipation and the transient response of any arbitrary phase variable after an external field is activated. TTCF has been applied to investigate the rheology of a range of fluids at low shear rates, usually in homogenous systems without confining walls\cite{doi:10.1080/08927020601026629}. These studies have progressed from atomic fluids\cite{PhysRevA.35.792,PhysRevA.38.4142,doi:10.1080/00268970210137275} to molecular fluids\cite{doi:10.1080/00268970902922625}, and even liquid metals\cite{PhysRevB.78.184202,doi:10.1063/1.2829872}. In addition to the viscosity, TTCF has also been used to monitor the electrical conductivity\cite{doi:10.1063/1.2035085,English2010,doi:10.1080/00268976.2010.544263}, thermal conductivity\cite{doi:10.1080/08927020801930604}, colour conductivity\cite{PhysRevE.77.027701}, and normal stress differences\cite{doi:10.1080/08927020802575598}. The TTCF method has also been extended from shear flow to elongational flow\cite{PhysRevE.56.6723,PhysRevE.58.4587} as well as mixed shear and elongational flow\cite{doi:10.1063/1.3684753}.
Relatively fewer studies have applied TTCF to boundary-driven simulations of confined, inhomogeneous systems\cite{PhysRevB.72.172201,22920110,doi:10.1080/08927022.2015.1049174}. 

Boundary-driven NEMD was introduced by Ashurst and Hoover\cite{PhysRevA.11.658} and further developed by Bitsanis et al. \cite{doi:10.1063/1.453240} who included the confining walls. In boundary-driven confined NEMD simulations, shear is usually applied by moving the walls in opposite directions and the temperature inside the channel is controlled by a thermostat applied either on the whole system or only to the wall atoms, as occurs in experiments\cite{PhysRevA.45.3706,doi:10.1063/1.3450302,Yong2013}. This approach enables the study of new dynamic behavior which emerges when fluids are strongly confined\cite{doi:10.1126/science.253.5026.1374,doi:10.1126/science.250.4982.792,doi:10.1063/1.473692,Xu6560} or subjected to very high pressure\cite{doi:10.1063/1.3698601,PhysRevE.88.052406,27802615}. Delhommelle and Cummings\cite{PhysRevB.72.172201} used TTCF to study the frictional response of a Weeks-Chandler-Anderson (WCA)\cite{doi:10.1063/1.1674820} fluid confined to a film of about five molecular diameters over a wide range of shear rates ($10^3-10^{11}\;\text{s}^{-1}$). However, the wall atoms lacked thermal motion and the fluid atoms were thermostatted, which makes an analytical derivation of the dissipation function difficult. Bernardi et al. used TTCF to study friction in WCA and Lennard-Jones (LJ) fluids confined between wall atoms with thermal motion\cite{22920110}. This meant that there was no ambiguity in the definition of the dissipation, which was analytically derived following its mathematical definition\cite{doi:10.1063/1.2812241}. More recently, Bernardi et al.\cite{doi:10.1080/08927022.2015.1049174} used TTCF to study the low shear rate rheology of confined polymer chains of several lengths (1, 2, 4, 8 and 12 beads) represented by the finite extensible nonlinear elastic (FENE) potential\cite{doi:10.1063/1.458541}.

In all previous TTCF studies of confined systems\cite{PhysRevB.72.172201,22920110,doi:10.1080/08927022.2015.1049174} fixed channel widths have been employed with a constant fluid volume. Experimentally, the channel width can often vary dynamically in response to pressure, shear, and temperature. Certain important behaviour, such as shear dilatancy\cite{26039993}, can only be captured in confined NEMD simulations if the channel width can change\cite{Xu6560}. This can be achieved by applying a barostat to the confining walls\cite{doi:10.1063/1.459524,PhysRevE.90.043302}. Quantitatively different flow and friction behaviour has been observed for systems at constant pressure rather than fixed channel width, i.e. constant volume\cite{PhysRevE.90.043302}. 
In this study, we use TTCF to investigate the friction of confined LJ and WCA fluids at low shear rates under the conditions of both constant volume and pressure. The temperature and pressure are controlled using a thermostat and barostat acting only on the wall atoms to closely mimic experimental conditions. In what follows, we first outline our methodology and establish the governing equations of motion and their associated TTCF expressions. Next, we present results for our TTCF computations for the shear pressure (negative of the shear stress) and slip velocity and compare them with their standard direct-average NEMD values. We end with some concluding remarks and present the TTCF derivation in the Appendix.

% METHODOLOGY
\section{Methodology and computation details}
\begin{figure}
	\centering
	
	\includegraphics[scale=0.4]{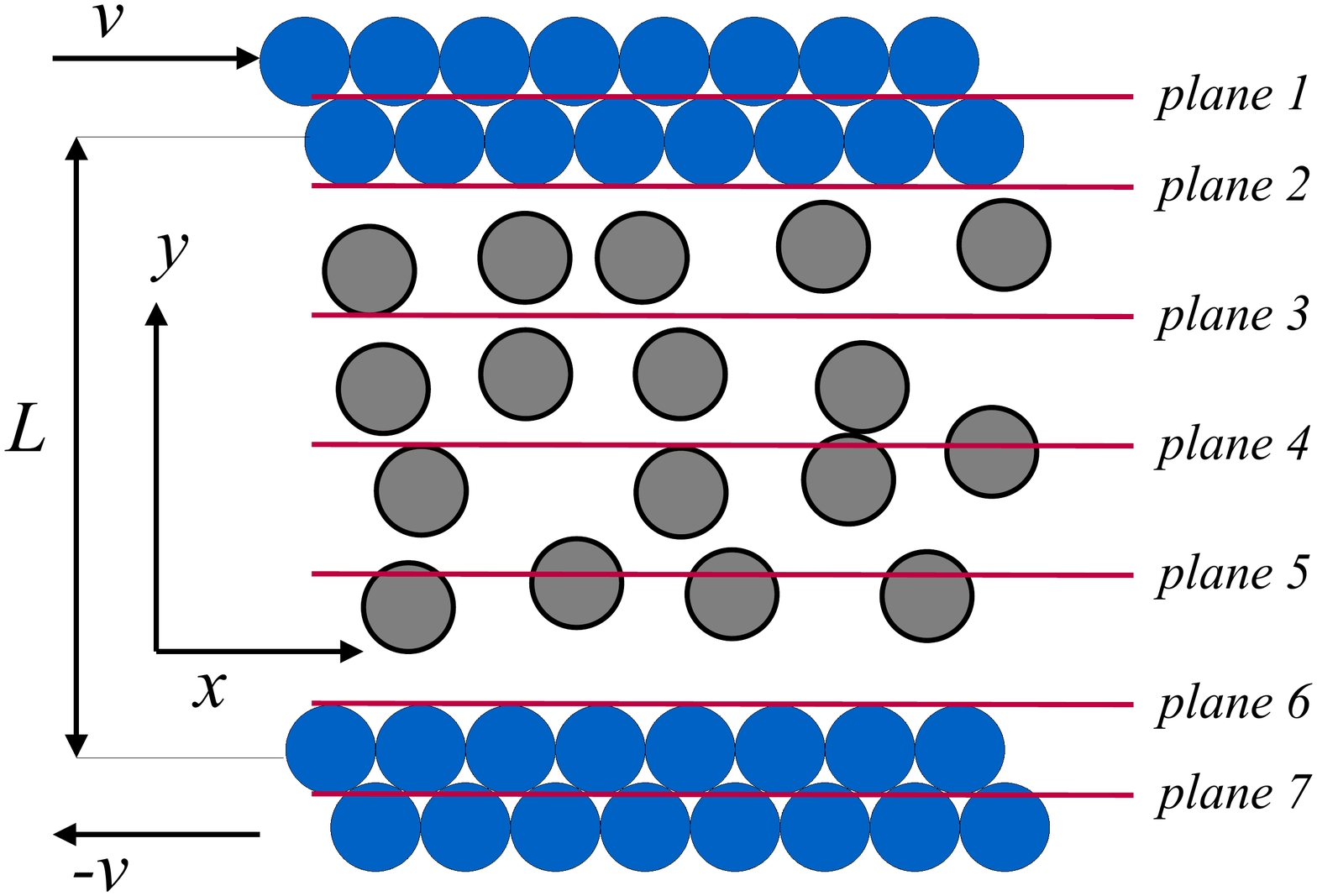}
	
	\caption{Schematic of the system and location of the planes across the channel. The $z$-direction is normal to the page. The diagram is merely representative and the size of the particles may not be proportionate to the actual system size.}
	\label{fig:system}
\end{figure}

\begin{figure}
	\centering
	
	\includegraphics[width=0.5\linewidth]{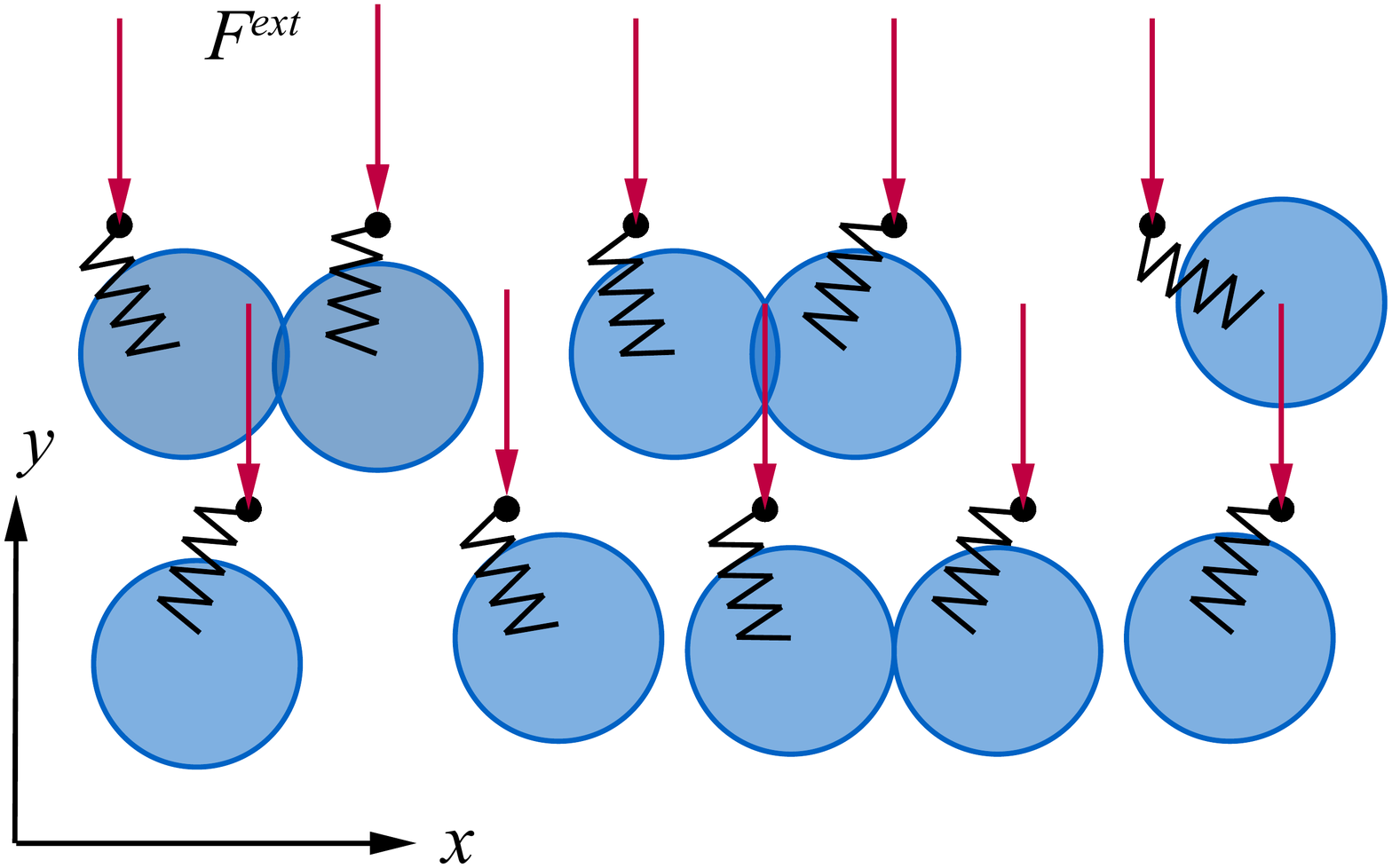}
	
	\caption{Schematic of the external field acting on the lattice sites (black dots). The lattice is subjected to both the tethering force (only some wall particles are actually shown, for the sake of clarity) and the external force. The action of the force occurs only along the $y$ (vertical) direction, and it does not affect the dynamics in the $x$ or $z$-directions.}
	\label{fig:barostat}
\end{figure}

%Transport coefficients such as the viscosity are often calculated from time-correlation functions\cite{doi:10.1146/annurev.pc.16.100165.000435}. TTCF generalises the popular Green-Kubo method\cite{doi:10.1063/1.1740082,doi:10.1143/JPSJ.12.570} which is strictly only applicable to equilibrium systems, to systems forced out of equilibrium by the imposition of an external field. For weak fields (i.e. low shear rates), TTCF offers far lower statistical uncertainty compared to direct NEMD\cite{PhysRevA.35.792,PhysRevA.38.4142}.
%In conventional NEMD simulations, the response is calculated using direct averaging (DAV), the dynamics of the system is followed over a single trajectory and an average is produced over N measurements. However, to generate acceptable statistics, i.e. a good signal  to noise ratio, high force fields must be applied\cite{22920110}. 
We firstly note that all of our simulations were performed using our own in-house code. Lupkowski and van Swol\cite{doi:10.1063/1.459524} invented a boundary-controlled barostat for controlling the normal pressure in molecular dynamics simulations of liquids confined between planar fluctuating walls, where a wall is represented by a solid block with one degree of freedom. Extensions of this barostat were derived by Gattinoni et al.\cite{PhysRevE.90.043302} who incorporated an atomistic description of the wall and where the atoms are tethered to an underlying `virtual’ rigid lattice which plays a similar role to the Lupkowski and van Swol barostat wall. In this work we adopt a mixed approach, where for our 3-dimensional system the wall particles are bound to a set of virtual lattice sites free to move along the $y$-direction (the confinement direction) and on which an external constant force is applied. The shear is likewise exerted on the system by applying a constant velocity to the lattice sites along the $x$-direction. The geometry of the system is depicted in Fig. \ref{fig:system} and the resulting equations of motion are:
\begin{equation}
\label{eqn:wall}
\begin{split}
&\text{constant volume}\\
\dot{\textbf{r}}^f_i&=\dfrac{\textbf{p}^f_i}{m_i}\\
\dot{\textbf{p}}^f_i&=\textbf{F}^{2B}_i\\
\dot{\textbf{r}}^w_i&=\dfrac{\textbf{p}^w_i}{m_i}\\
\dot{\textbf{p}}^w_i&=\textbf{F}^{2B}_i-k(\textbf{r}^w_i-\textbf{r}^l_i)-\alpha\textbf{p}^w_i\\
\dot{\alpha}&=\dfrac{1}{Q}\biggl(\sum_i^{N^w}\textbf{p}^{w2}_i-3N^wk_BT\biggr)\\
\dot{\textbf{r}}^l_i&=\bigl(\pm v\;\;,\;\;0\;\;,\;\;0\bigr)\\
\dot{\textbf{p}}^l_i&=\bigl(0\;\;,\;\;0\;\;,\;\;0\bigr).
\end{split}
\quad \quad \quad
\begin{split}
&\text{constant pressure}\\
\dot{\textbf{r}}^f_i&=\dfrac{\textbf{p}^f_i}{m_i}\\
\dot{\textbf{p}}^f_i&=\textbf{F}^{2B}_i\\
\dot{\textbf{r}}^w_i&=\dfrac{\textbf{p}^w_i}{m_i}\\
\dot{\textbf{p}}^w_i&=\textbf{F}^{2B}_i-k(\textbf{r}^w_i-\textbf{r}^l_i)-\alpha\textbf{p}^w_i\\
\dot{\alpha}&=\dfrac{1}{Q}\biggl(\sum_i^{N^w}\textbf{p}^{w2}_i-3N^wk_BT\biggr)\\
\dot{\textbf{r}}^l_i&=\bigl(\pm v\;,\;\dfrac{p_y^l}{m_i}\;,\;0\bigr)\\
\dot{\textbf{p}}^l_i&=\bigl(0\;,\;-F^{ext}+\dfrac{1}{N^l}\sum_i^{N^l}k(r_{yi}^w-r_{yi}^l)\;,\;0\bigr).
\end{split}
\end{equation}  
Here, the superscripts $f$, $w$ and $l$ denote, respectively, the fluid, wall, and lattice particles. $\textbf{F}_i^{2B}=-\sum_{i\ne j}\nabla \phi_{ij}$ is the interatomic two-body force, with
\begin{equation}
\phi\left(r_{ij}\right)= 
\begin{cases}
4\epsilon\biggl[\biggl(\dfrac{\sigma}{r_{ij}}\biggr)^{12}-\biggl(\dfrac{\sigma}{r_{ij}}\biggr)^6\biggr] + \phi_c,& \text{if } r_{ij}\leq r_c\\
0,              & \text{if } r_{ij}> r_c
\end{cases} 
\end{equation}
and $\textbf{F}_i^{H}=-k(\textbf{r}_i^w-\textbf{r}_i^l)$ is the harmonic force tethering the wall particles and the lattice sites (see Fig. \ref{fig:barostat}), while $\sigma$ and $\epsilon$ are the effective diameter and potential well, respectively. The cut-off radius is $r_c=2^{1/6}\sigma$ for Weeks-Chandler-Andersen\cite{doi:10.1063/1.1674820} (WCA) particles and $r_c=2.2\sigma$ for Lennard-Jones\cite{doi:10.1098/rspa.1924.0081} (LJ) ones. The term $\phi_c$ has the purpose to eliminate the discontinuity in the potential at $r=r_c$ and $\phi_c=\epsilon$ for the WCA potential and $\phi_c\simeq0.34\epsilon$ for the LJ. As will be clear from the following results, systems composed of simple atomistic WCA particles hardly exhibit any slip velocity at the wall-fluid interface. In order to provide a useful example of the TTCF formalism in the study of slip velocity, we investigated an analogous systems with particles interacting with a Lennard-Jones potential and where the parameters $\sigma$ and $\epsilon$ have been specifically modulated to promote a slip at the fluid-wall interface. In what follows, all quantities are expressed in reduced units, with $\sigma = \epsilon = 1$ for the WCA systems, and $\sigma^f=1$, $\sigma^w=0.5$, $\epsilon^f=\epsilon^w=1$ for the LJ one. In the latter, the interaction between fluid and wall particles has been modelled via the Lorentz-Berthelot mixing rule. For the case of constant volume, the lattices sites are fixed, and an initial velocity $\pm v$ (positive for the upper wall, negative for the lower one) is imposed in the $x$-direction. For the systems at constant pressure, the sites are free to move along the $y$-direction and are subjected to the mean harmonic force and to an additional external constant force $F^{ext}$ applied homogeneously to each site, which must be set to match the desired pressure. Since each site is subjected to the same force, the lattice sites move collectively as a single rigid body.

% fig 1

%Irving and Kirkwood (IK) derived formulas for the pressure tensor at a point in space.56 A number of alternative ‘local’ pressure tensor formulae have been derived since then. Here we obtain the local pressure tensor using the method of planes (MOP)57 which has shown to be mathematically identical58 to alternatives such as volume averaging (VA).59 In figure AAAA the position at which the pressure is computed is shown. 

For any generic phase variable $B\left(t\right)$, the TTCF formalism is based on the following identity\cite{doi:10.1063/1.2812241,22920110}:
\begin{equation}
\label{eqn:TTCF}
\langle B(t)\rangle=\langle B(0)\rangle+\int_0^t\langle\Omega(0)B(s)\rangle\text{d}s
\end{equation}
which relates the phase space average of $B$ at time $t$ with the time integral of the correlation with the dissipation function $\Omega(0)=\beta \dot{H}^{ad}$, i.e. the rate of energy dissipation without accounting for the thermostat term at $t=0$, when the external force is switched on\cite{doi:10.1063/1.3684753}. For our system we find that  
$\Omega=-\beta\sum_ik(r_{xi}^w-r_{xi}^l)v$ for both the constant volume and the constant pressure case.  See the Appendix for a brief derivation of the dissipation function in the presence of a barostat (the constant volume derivation results in the same expression and its derivation is analogous).

% fig 2

In summary, for the application of the TTCF method in computer simulations, one must generate a number $N_s$ of uncorrelated equilibrium ($v=0$) phase state points, which act as initial conditions of $N_s$ nonequilibrium systems ($v\ne0$) on which  $B(t)$ is computed. The multiple time series are then averaged over the $N_s$ measurements and integrated as per Eq. \eqref{eqn:TTCF}. To improve the statistics, it is useful to generate several different nonequilibrium trajectories from the same starting point, by generating transformed, or mapped, initial conditions from each starting point. In this work, we adopt four sets of phase-space mappings: for each particle $i$ the mappings are generated by the following rules, taken from previous works:\cite{PhysRevB.72.172201}
\begin{equation}
\begin{split}
\bigl(x_i\;,\;y_i\;,\;z_i\;,\;p_{xi}\;,\;p_{yi}\;,\;p_{zi}\bigr)&\longrightarrow\bigl(x_i\;,\;y_i\;,\;z_i\;,\;-p_{xi}\;,\;-p_{yi}\;,\;-p_{zi}\bigr)\\
\bigl(x_i\;,\;y_i\;,\;z_i\;,\;p_{xi}\;,\;p_{yi}\;,\;p_{zi}\bigr)&\longrightarrow\bigl(-x_i\;,\;y_i\;,\;z_i\;,\;-p_{xi}\;,\;p_{yi}\;,\;p_{zi}\bigr)\\
\bigl(x_i\;,\;y_i\;,\;z_i\;,\;p_{xi}\;,\;p_{yi}\;,\;p_{zi}\bigr)&\longrightarrow\bigl(-x_i\;,\;y_i\;,\;z_i\;,\;p_{xi}\;,\;-p_{yi}\;,\;-p_{zi}\bigr).\\
\end{split}
\end{equation} 
It is straightforward to verify that the above transformations are compatible with the canonical ensemble, which is symmetric by any of those transformations. This procedure ensures that $\langle\Omega(0)\rangle$ is identically $0$, as expected theoretically, and thus the integral of the time correlation will converge. Further details of the TTCF methodology and the use of phase-space mappings can be found in the specialist books on nonequilibrium statistical mechanics and nonequilibrium molecular dynamics \cite{EvansMorriss2008, ToddDaivis2017}.

We computed the $P_{xy}$ component of the local pressure tensor (the negative of the shear stress, or shear pressure) using the method of planes (MoP) technique\cite{PhysRevE.52.1627} at different locations across the channel, as per Figure \ref{fig:system} for the WCA systems. For constant pressure simulations the planes themselves are instantaneously adjusted since the channel width can fluctuate. While the absolute plane positions can fluctuate in time, their relative positions with respect to channel width remains constant. In addition, we generated velocity and density profiles of both WCA and LJ cases. Since the dissipation function $\Omega$ is strongly localized in the wall region, the calculation of any of the TTCF quantities at various distances from the walls allowed us also to verify whether the correlation $\langle\Omega\left(0\right) B\left(t\right)\rangle$ decays when $B\left(t\right)$ is computed far from the wall region, potentially making the TTCF method less effective. 

NEMD simulations for WCA systems composed of $3200$ fluid particles and $800$ wall particles were performed. Each wall was composed of two layers. All systems had initial density $\rho=0.8442$, $L\simeq 16 \sigma = 16$ and were thermostatted with a Nose-Hoover thermostat \cite{Nose1984, Hoover1985} at temperature $T=1$ at the walls. For the case of constant pressure, we analysed systems at relatively high pressures of $P=10$ and $P=11$. Higher pressures than this might activate a phase transition where the fluid region becomes solid \cite{doi:10.1063/1.3698601, PhysRevE.88.052406, 27802615}, a condition for which the TTCF formalism fails or becomes highly inefficient. This phenomenon might also be promoted by a small system size along the $x$-direction. For the LJ case, we investigated systems at fluid density $\rho^f=0.8$ and $\rho^w=8$, and a channel width of $\simeq 10\sigma$. The high density of the wall is due to the reduced diameter of the wall particles, and corresponds to $\rho^w=1$ if the quantities were normalized assuming $\sigma^w=1$. A recent work\cite{PhysRevE.100.023101} pointed out that small inhomogeneous systems, under certain conditions,  tend to underestimate the solid-liquid friction. However, in this work we have not accounted for this effect, as we focussed solely on the derivation and application of the TTCF formalism under constant pressure conditions for the study of highly confined fluid systems. The reduced mass of each particle and lattice site was set to $1$, and the stiffness of the spring was set as $k=150$ for the WCA systems and $k=75$ for the LJ ones \cite{PhysRevE.55.4288}. 
  
Each system underwent an equilibration $400$ time units, and subsequently $10^5$ initial conditions were sampled, for a total $4\times10^5$ starting points for the TTCF calculation due to the mappings, with a delay of $2$ time units between each sample along the equilibrium simulations. Each nonequilibrium trajectory was followed for $10$ time units. Since the TTCF method is based on a time correlation at time $t=s$ and $t=0$, it is essential that the equations of motion are integrated with a self-starting integrator; this means that the popular Gear predictor-corrector would not be suitable for the purpose. A fourth order Runge-Kutta is typically used\cite{doi:10.1080/08927020801930604,doi:10.1080/08927020802575598} which, on the other hand, make the simulation computationally expensive. We found that the faster velocity Verlet performs equally well. The equations of motion of both the equilibrium and nonequilibrium systems were therefore integrated using a velocity Verlet algorithm \cite{Verlet1967}, with time step $\delta t=0.004$.

% RESULTS
\section{Results}
\begin{figure}
	\centering
	
	\includegraphics[scale=1.1]{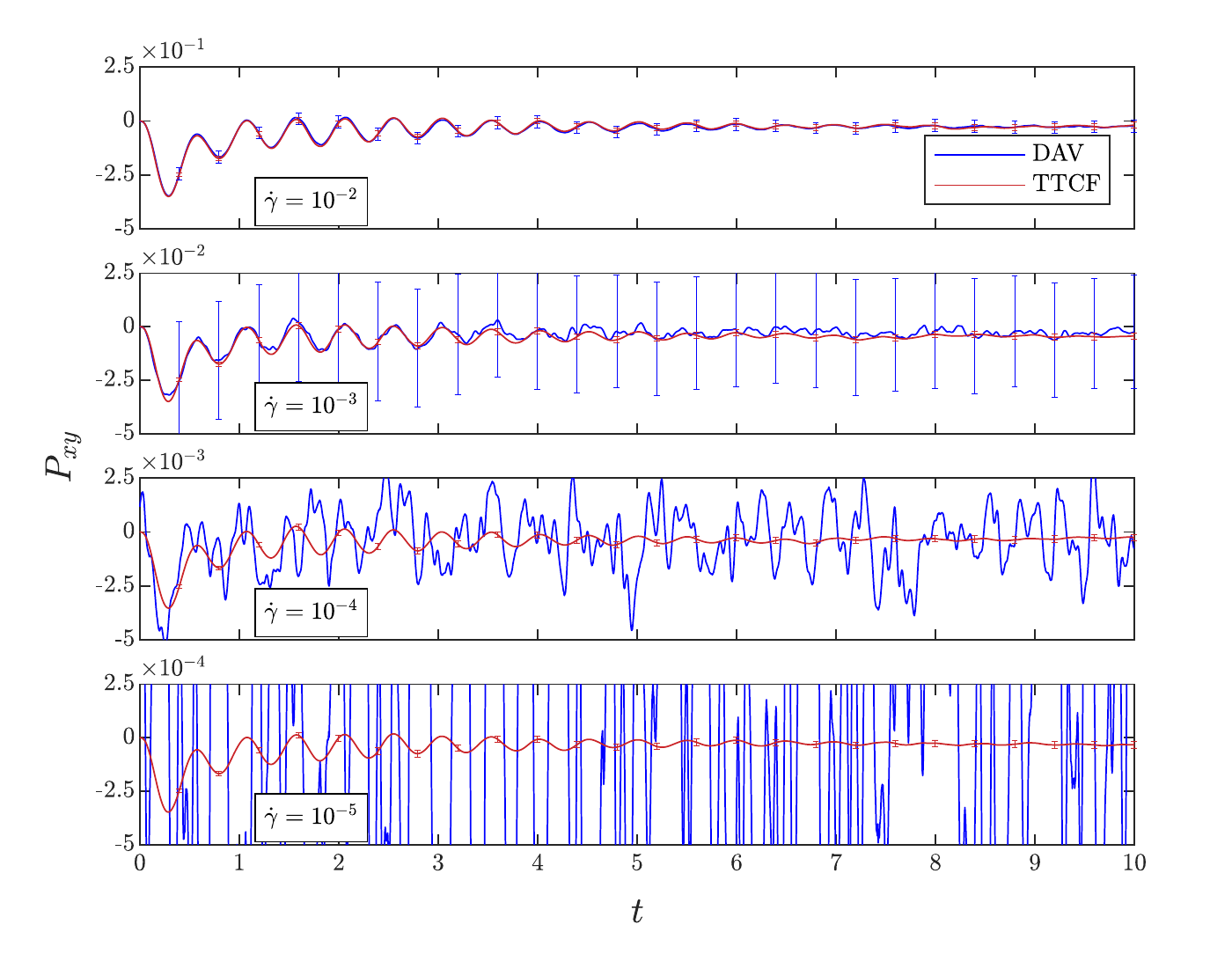}
	
	\caption{$P_{xy}$ at the fluid/wall interface, obtained by averaging over planes 2 and 6 (see Fig. \ref{fig:system}), for the WCA systems at constant volume and for different shear rates. The size of the error bars is 4 times the standard error. For the sake of clarity, the error bars of the DAV method for the two lowest shear rates are not displayed and are roughly an order of magnitude larger than the signal itself. A constant reduction of the uncertainty in the TTCF time series indicates that this method would be highly accurate for even lower shear rates.}
	\label{fig:pxyvsshear}
\end{figure}

Figure \ref{fig:pxyvsshear} shows the comparison between the direct average (DAV) and the TTCF methods in computing the shear pressure $P_{xy}\left(y\right)$ as a function of time for WCA systems at the fluid/wall interface, for the case of a constant volume system and reduced shear rates ranging between $\dot{\gamma}=10^{-2}-10^{-5}$, which correspond approximately to $\simeq5\times10^9-5\times10^6\;\text{s}^{-1}$ in MKS units, assuming that the fluid is composed of monatomic argon atoms \cite{doi:10.1063/1.479848}. The DAV data are obtained by simply averaging the time dependent relevant phase variable (in this case, the negative of the shear stress $\sigma_{xy}$, where $\sigma_{xy} \equiv -P_{xy}$) over multiple independent NEMD trajectories. The number of these trajectories is simply the number of initial conditions (i.e. $1 \times 10^5$), whereas the number of TTCF trajectories, as described in the previous section, is $4\times10^5$ trajectories, due to the phase-space mappings required. Even though the number of TTCF trajectories is four times more than the number of DAV trajectories, we do not average over all four phase-space mappings for the DAV data. The reason for this is that at short times the DAV trajectories are highly correlated and fluctuations in statistical error cancel due to mapping symmetry. However, Lyapunov instability leads to a rapid de-correlation as time increases. In the long-time limit (and certainly in the steady-state) the standard error would only be improved by a factor of two if we were to use all $4\times10^5$ trajectories for DAV. A factor of two is still insignificant compared to the improvement in statistics we find with the TTCF results, as we will demonstrate shortly. 

We observe at early times the negative shear stress overshoots\cite{Heyes1980}, followed by damping oscillatory behaviour, similar to that observed by Bernardi et. al \cite{22920110}, due to the atomic vibrations in wall atoms (Fig. \ref{fig:pxyvsshear}). For the highest shear rate, the statistical accuracy of the two methods is comparable. The data show that, for the systems analysed, the standard error of the signal $P_{xy}$ computed with the DAV method is in the order of $10^{-2}$ or higher, and roughly constant for each level of the external force. This means that for $\dot{\gamma}<10^{-2}$ the time evolution of $P_{xy}$ is either inaccurate or totally indistinguishable from random noise, as can be seen in the last two plots. On the other hand, the TTCF method produces an extremely accurate and clean signal for arbitrarily low shear rates, with a standard error $\epsilon_{std}$ decreasing with the shear rate and spanning from approximately $\epsilon_{std}\approx10^{-2}$ for $\dot{\gamma}=10^{-2}$ to $\epsilon_{std}\approx10^{-5}$ for $\dot{\gamma}=10^{-5}$. The constant reduction of the uncertainty in $P_{xy}$ computed with the TTCF method indicates that an exceptionally good signal-to-noise ratio can be achieved at shear rates lower that those adopted in this work.   

% fig 3

\begin{figure}
	\centering
	
	\includegraphics[scale=0.9]{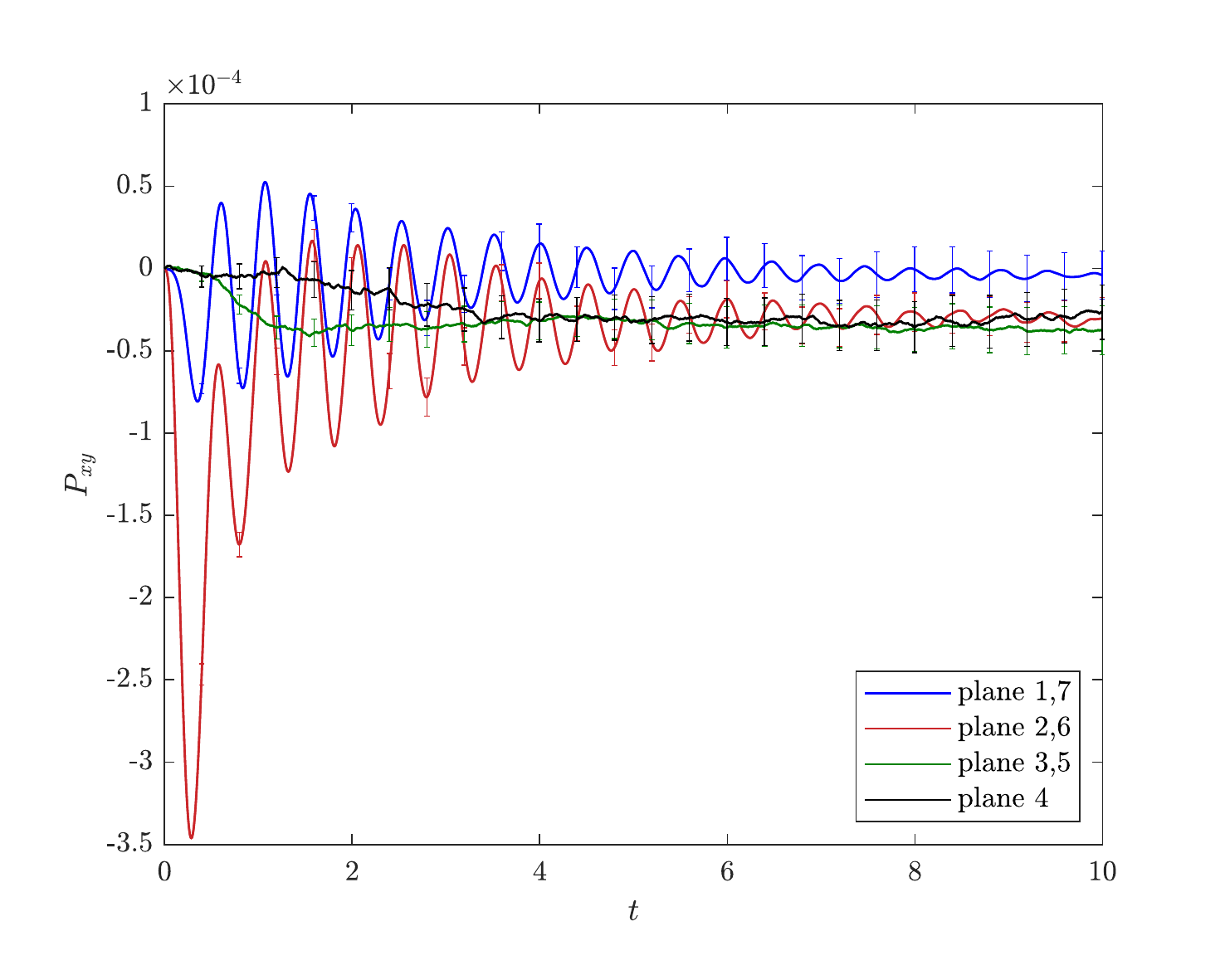}
	
	\caption{$P_{xy}$ at various locations in the system (see Fig. \ref{fig:system}) under constant volume for $\dot{\gamma}=10^{-5}$, computed with the TTCF formalism. The size of the error bars is 4 times the standard error.}
	\label{fig:pxyvsplane}
\end{figure}

In Figure \ref{fig:pxyvsplane} we see $P_{xy}$ at the various locations across the channel for $\dot{\gamma}=10^{-5}$, computed with the TTCF method under conditions of constant volume. The shear pressure in the fluid region (planes 3, 5 and 4) converges to a constant value across the channel, as expected theoretically. However, the transient region ($t < 6$) shows little or no oscillatory behaviour found close to the wall. In the wall region (planes 1, 7) the average shear pressure is essentially null, due to the constant drag of the lattice sites; however, the oscillatory behaviour due to the vibrating wall atoms is still apparent. The signal is particularly stable inside the walls and at the wall-fluid interface (planes 2, 6), due to the absence of the kinetic contribution, which is inherently noisier. Each curve retains similar levels of uncertainty. This indicates that a strong correlation between $\Omega\left(0\right)$ and the fluid evolution holds arbitrarily far from the walls. We also note that the steady-state shear pressure in the fluid region close to the wall (planes 3, 5) is reached more quickly than that near the channel centre (plane 4) due to the time it takes for momentum transfer from the walls into the bulk of the fluid.

% fig 4

\begin{figure}
\begin{subfigure}{.5\textwidth}
	\centering
	\includegraphics[scale=0.6]{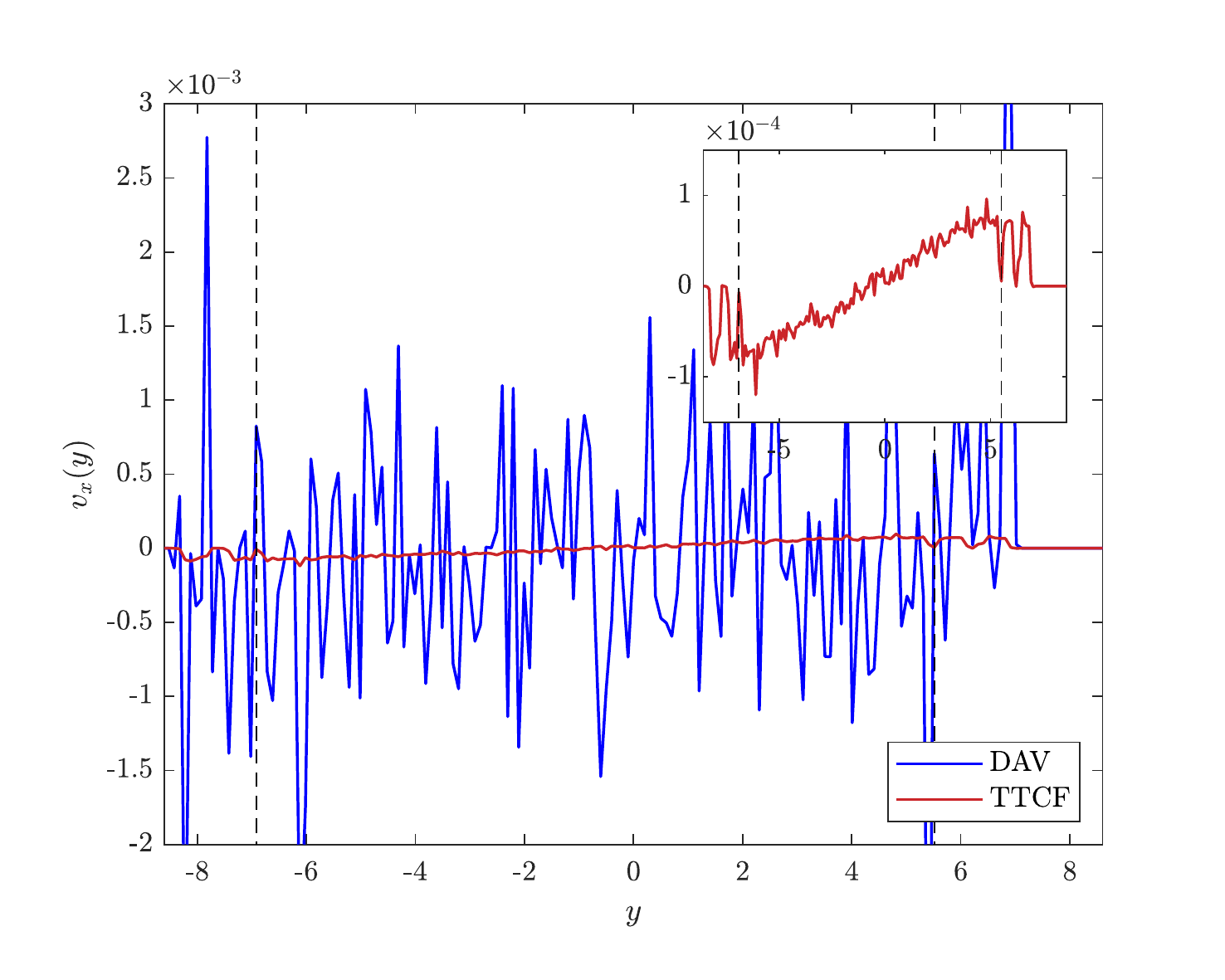}
	\caption{}
	\label{fig:velvol}
\end{subfigure}%
\begin{subfigure}{.5\textwidth}
	\centering
	\includegraphics[scale=0.6]{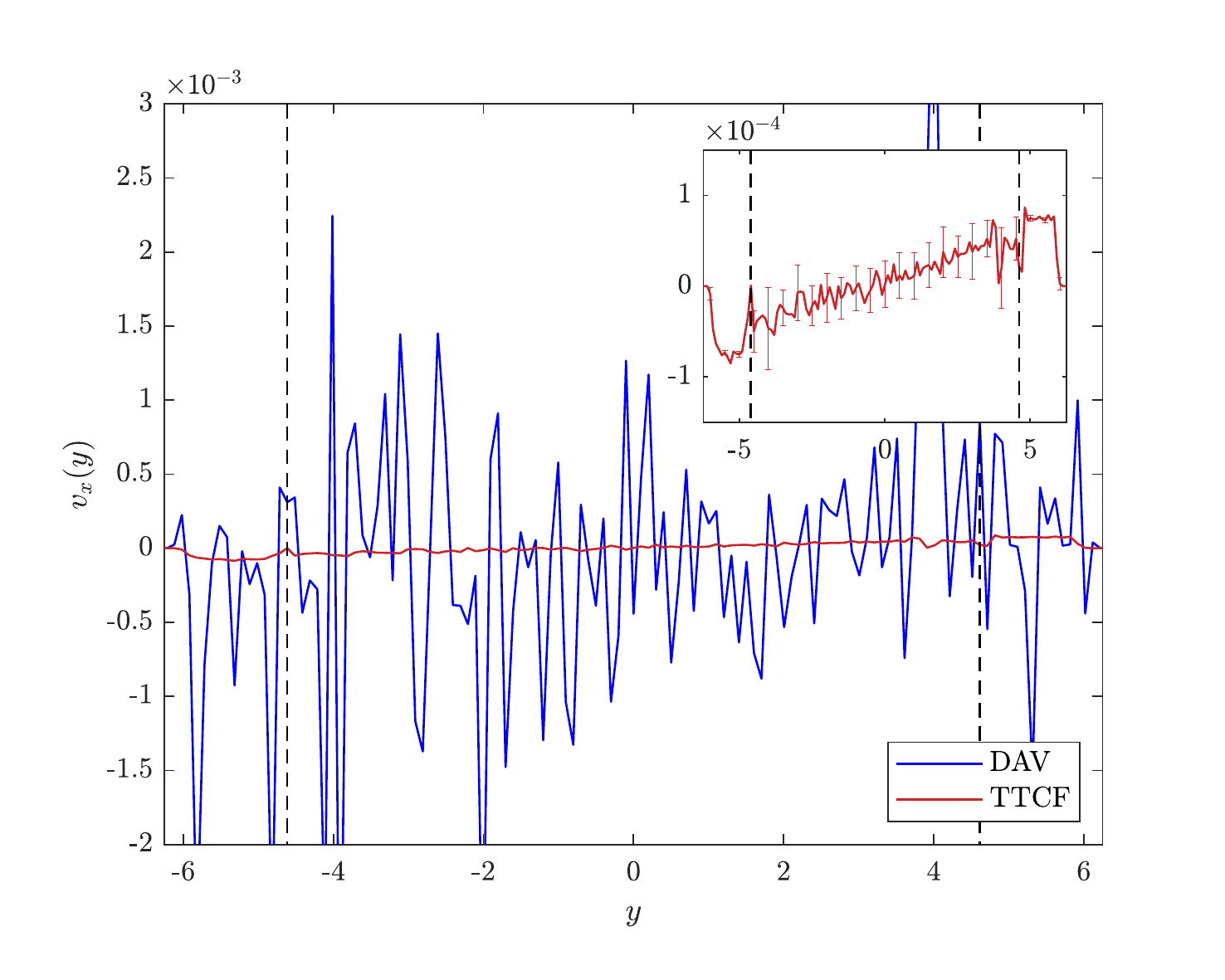}
	\caption{}
	\label{fig:velpres}
\end{subfigure}
	\caption{Velocity profile of the WCA system at constant pressure (left) at $t=10$ and LJ system (right) at $t=7.6$, for $\dot{\gamma}=10^{-5}$. The wall region is included, and can be recognised from the structured peaks in the velocity at the borders of the region. The vertical black dashed lines indicate the positions of the wall-fluid interface. The standard error of the DAV method has been omitted and it has a magnitude in the range of $10^{-2}$. Note the clearly defined linear velocity profile observed in the TTCF signal, as opposed to the direct average signal, which is too noisy to allow observation of any meaningful profile. The time $t=7.6$ for the LJ system has been chosen to show the slip in the velocity profile (compare with Figure \ref{fig:velslip2}).}
	\label{fig:velprofile}
\end{figure}

Figures \ref{fig:velvol} and \ref{fig:velpres} show the comparison between the DAV and TTCF methods in the computation of the streaming velocity profile at low shear rates ($\dot{\gamma}=10^{-5}$) for the systems at constant volume and constant pressure, defined as
\begin{eqnarray}
\mathbf{v}\left(\mathbf{r}_{\text{bin}}, t\right) = \frac{\langle \sum_{i\in{\text{bin}}} m_i \mathbf{v}_i \rangle}{\langle \sum_{i\in{\text{bin}}} m_i \rangle},
\end{eqnarray}
where the sum ranges over all fluid atoms within a bin of finite width with mid-point located at $\mathbf{r}_{\text{bin}}$. The standard error of the velocity computed via DAV is in the order of $10^{-2}$ which is more than two orders of magnitude larger than the signal itself. The velocity profile is critical in the calculation of the slip velocity and the corresponding slip length. In confined boundary driven systems on smooth surfaces, the velocity profile of the fluid deviates from the linear, no-slip velocity profile imposed by the motion of the walls. In particular, the average velocity of the fluid particles in proximity of the wall is smaller than that of the wall and lower than the theoretical one \cite{PhysRevA.41.6830}. Of particular significance in these results is that TTCF has for the first time been used to determine accurate velocity profiles for a confined fluid under very low (from an NEMD simulation perspective) rates of strain that approach those values achievable under laboratory conditions \cite{Neto_2005}. The direct average (DAV) of the signal is simply too noisy and no information can be gleaned. However, the TTCF velocity profiles, as seen from the insets in Figs. \ref{fig:velvol} and \ref{fig:velpres}, demonstrate that the signal can be clearly distinguished from the noise. This in itself is a significant achievement and has never been seen before. It finally opens the way for the use of NEMD to precisely compute stresses, velocities or any other phase variable whatsoever, at physically meaningful and accessible laboratory strain rates.

% fig 5

\begin{figure}
	\begin{subfigure}{.5\textwidth}
		\centering
		\includegraphics[scale=0.6]{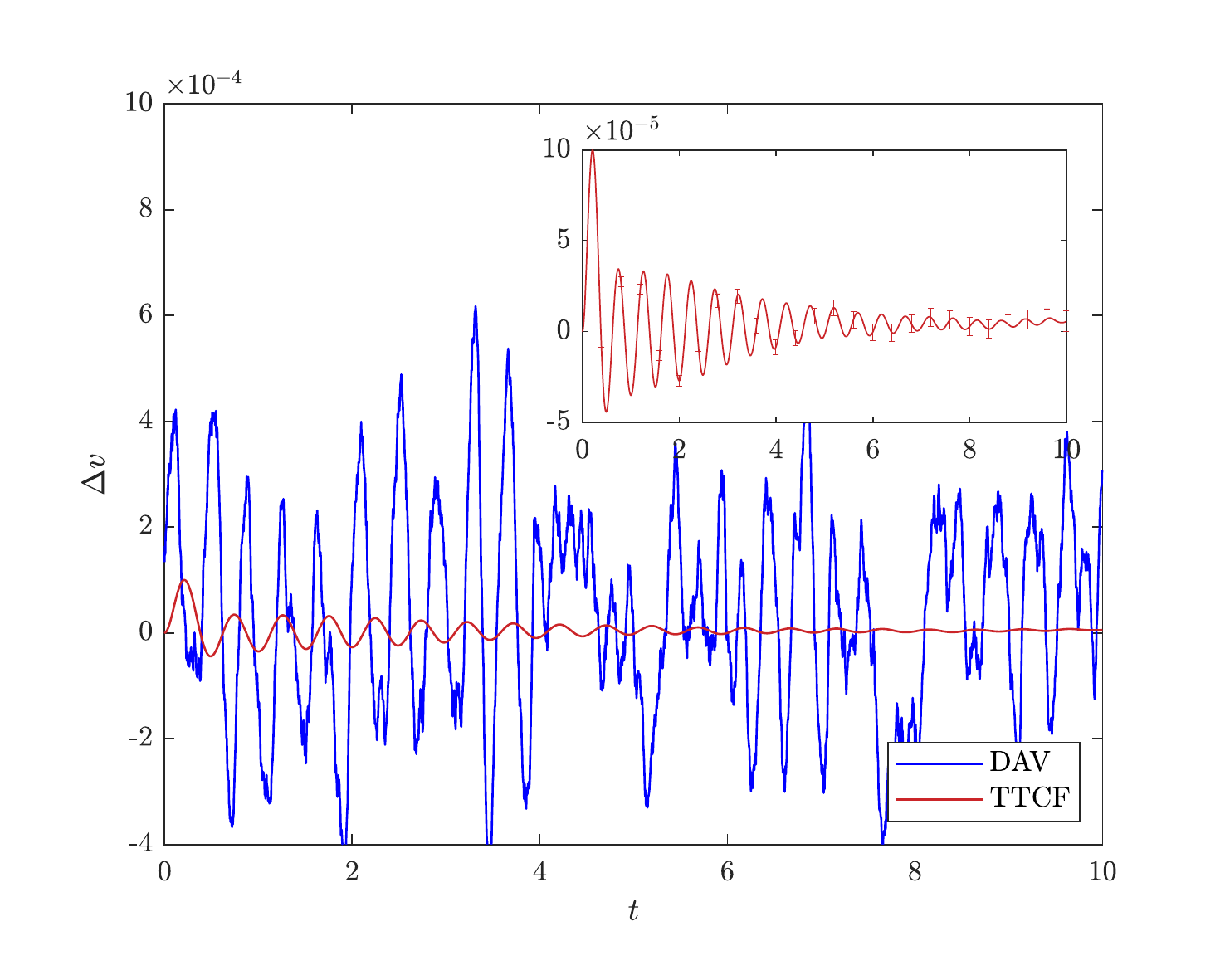}
		\caption{}
		\label{fig:velslipvol}
	\end{subfigure}%
	\begin{subfigure}{.5\textwidth}
		\centering
		\includegraphics[scale=0.6]{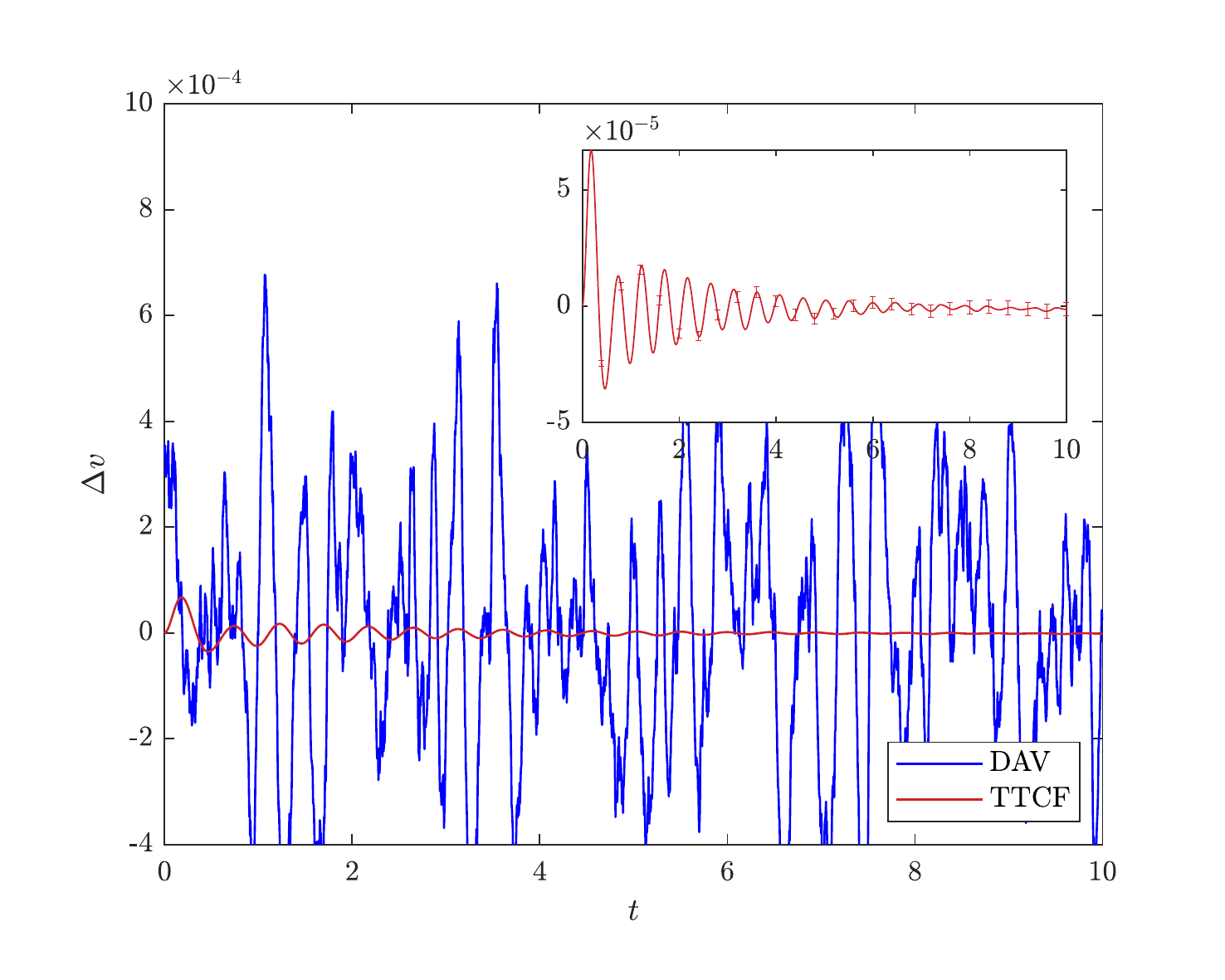}
		\caption{}
		\label{fig:velslippres}
	\end{subfigure}
	\caption{Slip velocity at the interface for the WCA system at constant volume (a) and pressure (b)  for $\dot{\gamma}=10^{-5}$. The error bars for the DAV measurements have been omitted since they are at least an order of magnitude larger than the signal.}
	\label{fig:velslip}
\end{figure}

\begin{figure}
	\begin{subfigure}{.5\textwidth}
		\centering
		\includegraphics[scale=0.6]{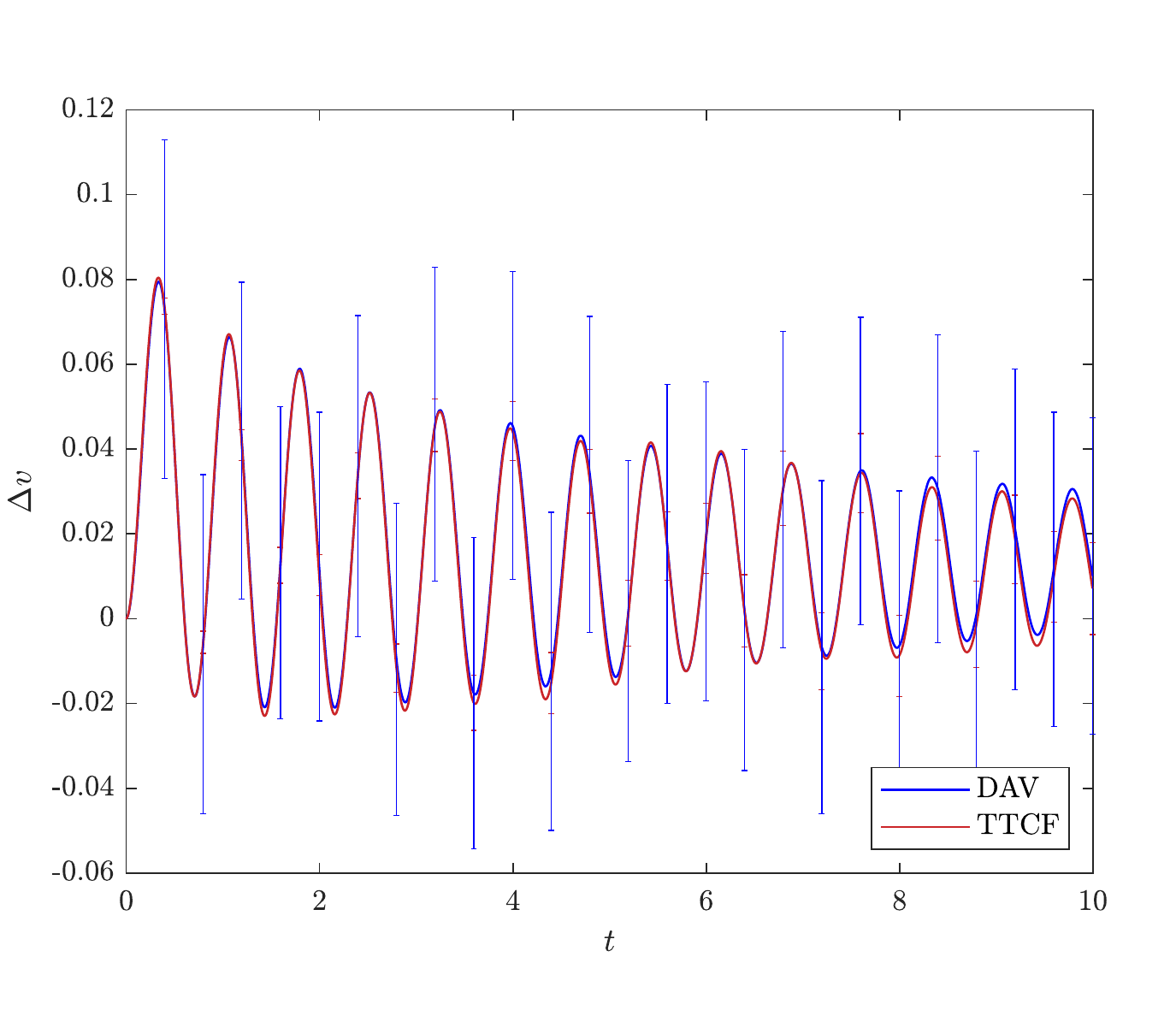}
		\caption{}
		\label{fig:velslipvol2}
	\end{subfigure}%
	\begin{subfigure}{.5\textwidth}
		\centering
		\includegraphics[scale=0.6]{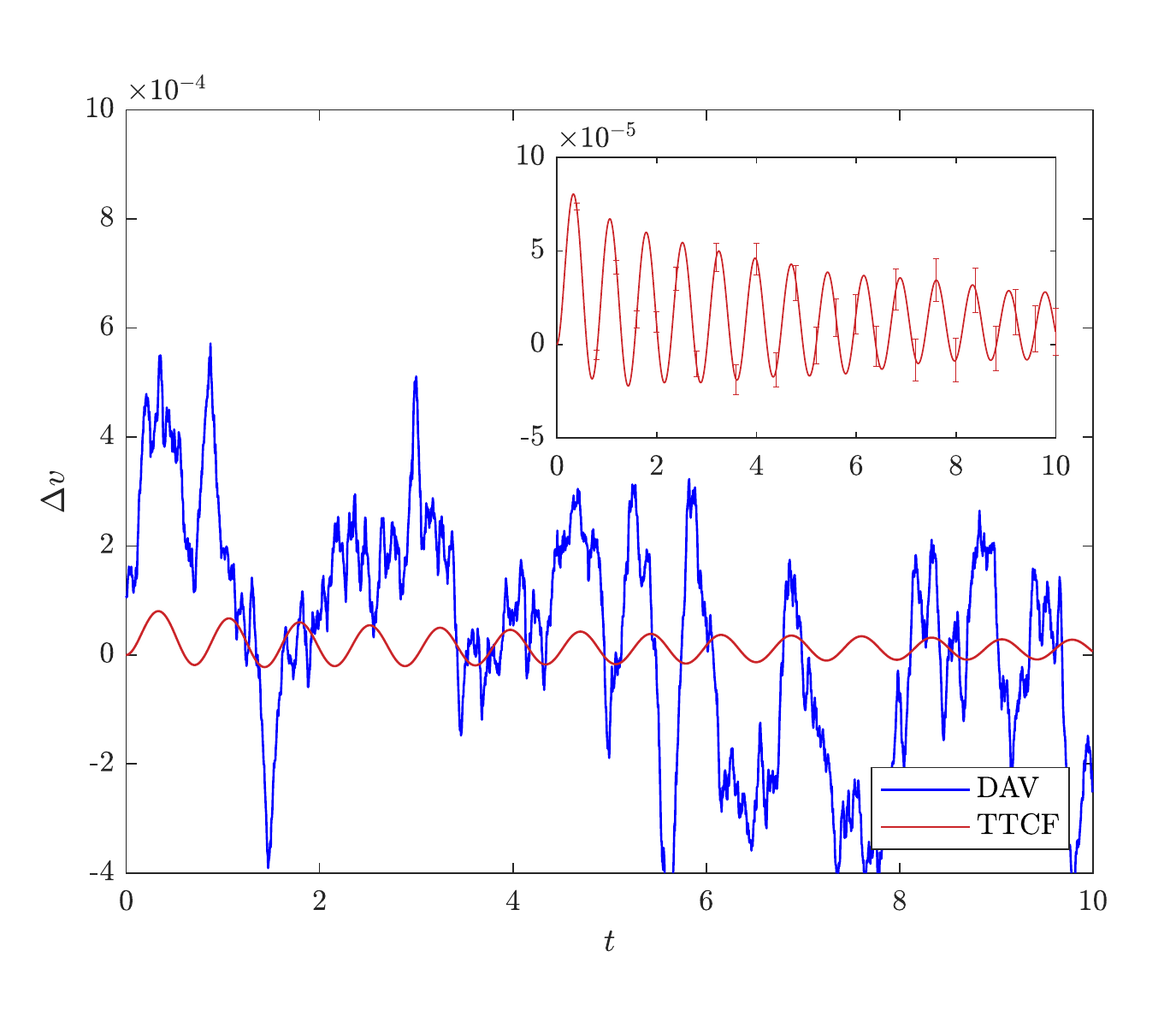}
		\caption{}
		\label{fig:velslippres2}
	\end{subfigure}
	\caption{Slip velocity at the interface at constant volume for $\dot{\gamma}=10^{-2}$ (a) and $\dot{\gamma}=10^{-5}$ (b) for the LJ system.}
	\label{fig:velslip2}
\end{figure}

Figures \ref{fig:velslip} and \ref{fig:velslip2} show more clearly the slip velocity, defined here as the difference in the velocity between the inner layer of the wall particles and the first layer of the fluid particles for WCA and LJ systems, respectively. In our systems, the layers have a width of roughly $\sigma$. The value has been taken as the average of the two interfaces, with the convention that the velocities are positive, i.e. a slip velocity greater than zero indicates that that streaming velocity of the wall is larger than that of the fluid. The data show again the power of the TTCF method. The computation of the slip velocity at low shear rates would be unfeasible if the direct average method was used, and it is typically difficult to estimate directly even at high shear rates, particularly for high-slip systems \cite{doi:10.1063/1.3675904}, due to the large fluctuations occurring in the velocity profile. It is precisely for such reasons that equilibrium molecular dynamics methods have been developed to compute the friction coefficients for such systems \cite{BB1994, HTD2011}, and have been shown to be more reliable than NEMD for systems of high slip, such as water confined to graphene or carbon nanotubes \cite{doi:10.1063/1.3675904,Kannam2013}. The TTCF results generated here now arm us with a powerful NEMD technique to obtain slip velocities, slip lengths and friction coefficients under actual nonequilibrium conditions. We further note that the oscillatory behaviour of the slip velocity is most likely due to the vibrations of wall atoms tethered to their lattice sites, rather than evidence of stick-slip behaviour such as that seen previously \cite{doi:10.1126/science.250.4982.792}. A more detailed analysis of this oscillatory behaviour will follow in a future publication.

\begin{figure}
	\centering
	
	\includegraphics[scale=0.9]{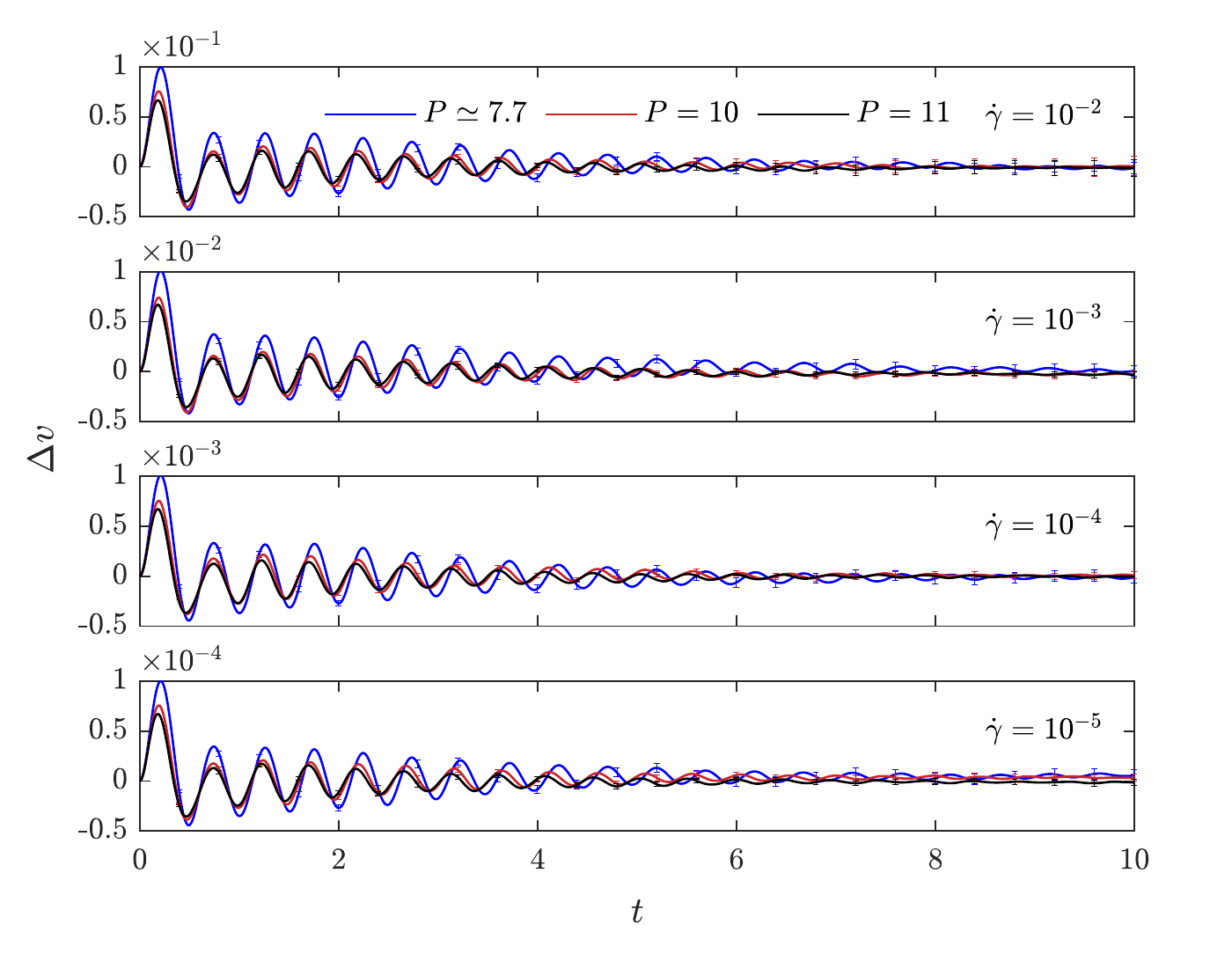}
	
	\caption{Slip velocity for all the systems investigated, computed with the TTCF method. $P\simeq7.7$ represents the systems at constant volume. There is no appreciable change in the behaviour of the systems at the interface in the various systems, which means that none of them show slip velocity. Error bars are 4 times the standard error.}
	\label{fig:velsliptable}
\end{figure}

\begin{figure}
	\centering
	
	\includegraphics[scale=0.9]{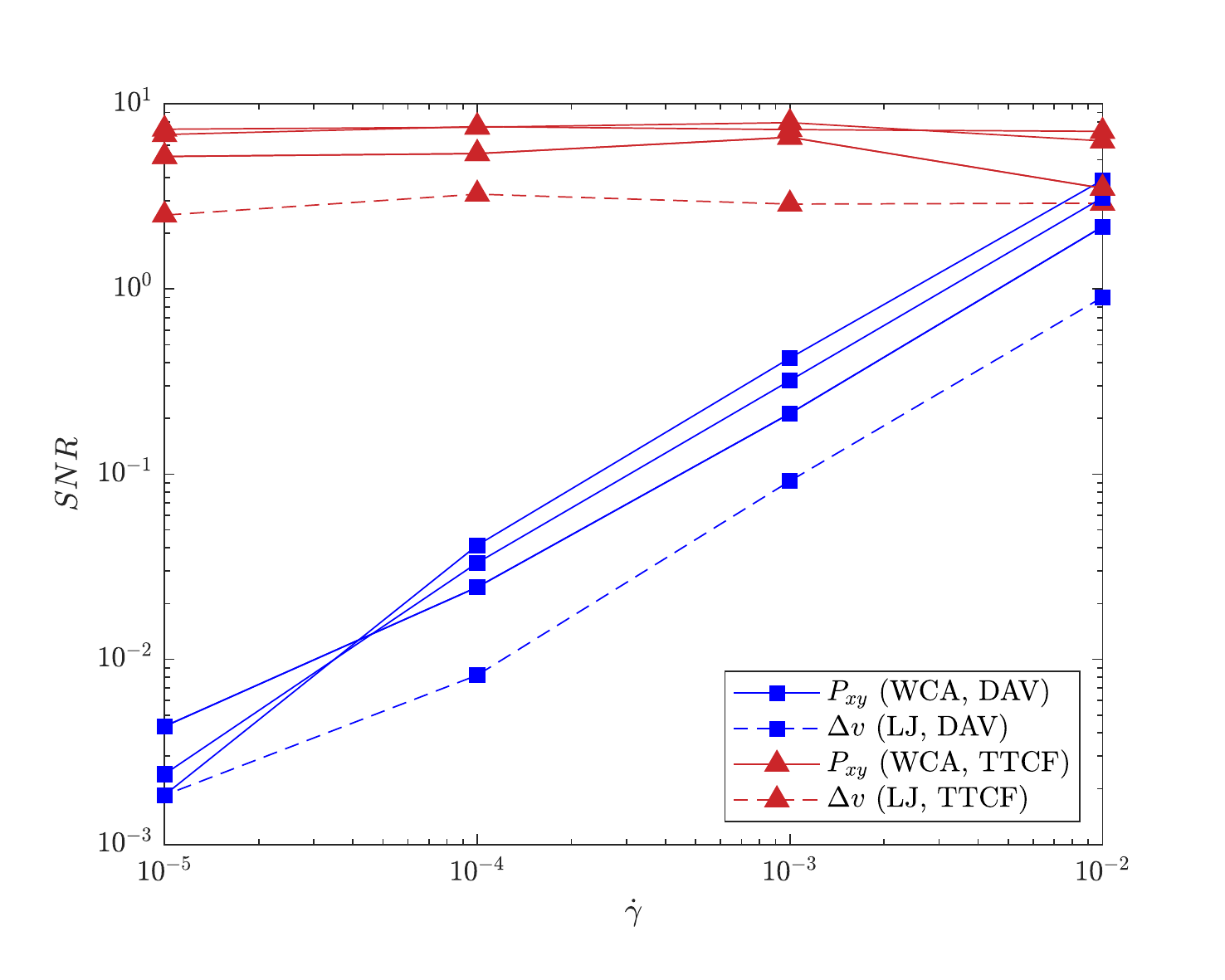}
	
	\caption{Signal-to-noise ratio for the shear pressure $P_{xy}$ of the WCA systems and for the slip velocity of the LJ system at the various shear rates. The results of the slip velocity for the WCA systems have been omitted since the signal itself is close to 0. The quality of the signal drops rapidly for low shear rates for the direct average, but remains unchanged for the TTCF.}
	\label{fig:snr}
\end{figure}

In Figure \ref{fig:velsliptable} the slip velocity, computed with the TTCF method for all the WCA systems, is displayed. The drop in the slip velocity is approximately proportional to the shear rate, and it is apparently not affected by the pressure imposed on the system except for the lowest shear rate. 
%Figure \ref{fig:velslipreltable} displays the ratio between the slip velocity and the theoretical one imposed by the shear, i.e. $\dot{\gamma}\delta r $ where $\delta r $ is the distance between the inner wall layer and the outermost fluid layer. The data are not significant in the first transient period, where the velocity profile deviates from linearity, but they indicate that there is essentially no slip when the steady state is reached. 
Tables \ref{tab:slipsummary1}, \ref {tab:slipsummary2} and \ref{tab:pressummary2} summarise the data. As can be seen, the error of the direct average rapidly dominates the signal for both the quantities measured. It is useful to compare the signal-to-noise ratio (SNR) for the DAV and the TTCF signals, defined as the ratio between the absolute value of the average of the signal and its standard error, at $t=10$. As can be seen in Figure \ref{fig:snr} the SNR remains constant for the TTCF method for each level of the external field. This is a clear and promising evidence that accurate measurements can be achieved with the same system size and number of samples for shear rates lower than studied in this work, approaching values which are practical.

% fig 7

% fig 8

% fig 10
%\begin{figure}
%	\centering
%	
%	\includegraphics[scale=0.9]{velsliptablerel}
%	
%	\caption{Slip velocity for all the systems investigated relative to the theoretical one, computed with the TTCF method. All the systems exhibit similar behaviour, and the slip velocity is essentially negligible, i.e. the is almost perfect stick between the wall and the first fluid layer. Error bars are 4 times the standard error. The slip velocity increases proportionally to the shear rate, but remains a tiny fraction of the theoretical one.}
%	\label{fig:velslipreltable}
%\end{figure}

% TABLE 1
\begin{table}
\caption{\label{tab:slipsummary1} Summary of the estimated slip velocity for the systems analysed. The system at $P\simeq7.7$ is the case at constant volume. Aside from the values itself, the superiority of the TTCF method can be appreciated by the reduced uncertainty in the final estimate. The $n$ values in brackets are twice the standard error and is the error in the last $n$ digits of the slip velocities quoted.}
	\begin{center}
		\begin{tabular}{| c| c| c| c| c| c| c| c |} 
			\hline
			\multicolumn{2}{|c|}{$P$} & \multicolumn{2}{c|}{$\simeq7.7$} & \multicolumn{2}{c|}{$10$} & \multicolumn{2}{c|}{$11$}\\ 
			\hline
			\multicolumn{2}{|c|}{$\dot{\gamma}$} & DAV & TTCF & DAV & TTCF & DAV & TTCF  \\
			\hline
			\multicolumn{2}{|c|}{$0.01$}    & 0.0018(44) & -0.0012(55) & 0.0005(49) & 0.0006(96) &-0.0004(47) & 0.0012(80)  \\
			\multicolumn{2}{|c|}{$0.001$}   & 0.0004(46) &  0.00008(5) & 0.0000(48) & -0.0003(4) & 0.0000(47) & -0.0003(3)  \\
			\multicolumn{2}{|c|}{$0.0001$}  &-0.0005(45) & -0.00001(6) & 0.0001(48) & 0.00001(3) & 0.0003(46) & -0.00000(2) \\
			\multicolumn{2}{|c|}{$0.00001$} & 0.0003(45) &  0.00000(0) & 0.0000(50) & 0.00000(0) & 0.0000(50) & 0.00000(0) \\
			\hline
		\end{tabular}
	\end{center}
%	\caption{\label{tab:slipsummary1} Summary of the estimated slip velocity for the systems analysed. The system at $P\simeq7.7$ is the case at constant volume. Aside from the values itself, the superiority of the TTCF method can be appreciated by the reduced uncertainty in the final estimate. The $n$ values in brackets are twice the standard error and is the error in the last $n$ digits of the slip velocities quoted.}
\end{table}

% TABLE 2
\begin{table}
	\caption{\label{tab:slipsummary2} Summary of the estimated slip velocity for the LJ system. Aside from the values itself, the superiority of the TTCF method can be appreciated by the reduced uncertainty in the final estimate. The $n$ values in brackets are twice the standard error and is the error in the last $n$ digits of the slip velocities quoted.}
	\begin{center}
		\begin{tabular}{|c | c| c| c| }  
			\hline
			\multicolumn{2}{|c|}{$\dot{\gamma}$} & DAV & TTCF \\
			\hline
			\multicolumn{2}{|c|}{$0.01$}    & 0.01688(3733) & 0.01584(1088) \\
			\multicolumn{2}{|c|}{$0.001$}   & 0.00175(3817) &  0.00172(120) \\
			\multicolumn{2}{|c|}{$0.0001$}  & 0.00015(3723) & 0.00019(12)   \\
			\multicolumn{2}{|c|}{$0.00001$} & 0.00004(3880) &  0.00001(1) \\
			\hline
		\end{tabular}
	\end{center}
%	\caption{\label{tab:slipsummary2} Summary of the estimated slip velocity for the LJ system. The system at $P\simeq7.7$ is the case at constant volume. Aside from the values itself, the superiority of the TTCF method can be appreciated by the reduced uncertainty in the final estimate. The $n$ values in brackets are twice the standard error and is the error in the last $n$ digits of the slip velocities quoted.}
\end{table}

% TABLE 3
\begin{table}
	\caption{\label{tab:pressummary2} Summary of the estimated shear pressure (negative shear stress) in the fluid region, as the average over the planes $2-6$. The system at $P\simeq7.7$ is the case at constant volume. The $n$ values in brackets are twice the standard error and is the error in the last $n$ digits of the shear pressures quoted.}
	\begin{center}
		\begin{tabular}{| c| c| c| c| c| c| c| c |} 
			\hline
			\multicolumn{2}{|c|}{$P$} & \multicolumn{2}{c|}{$\simeq7.7$} & \multicolumn{2}{c|}{$10$} & \multicolumn{2}{c|}{$11$}\\ 
			\hline
			\multicolumn{2}{|c|}{$\dot{\gamma}$} & DAV & TTCF & DAV & TTCF & DAV & TTCF  \\
			\hline
			\multicolumn{2}{|c|}{$0.01$}    &-0.0274(251) & -0.0205(117) &-0.0443(284) & -0.0476(134) &-0.0563(290) & -0.0475(151)  \\
			\multicolumn{2}{|c|}{$0.001$}   &-0.0027(259) & -0.0039(11) &-0.0045(283) & -0.0052(4) &-0.0057(268) & -0.0062(016)  \\
			\multicolumn{2}{|c|}{$0.0001$}  &-0.0003(247) & -0.0003(1) &-0.0004(283) & -0.0005(1) &-0.0006(286) & -0.0006(2) \\
			\multicolumn{2}{|c|}{$0.00001$} & 0.0000(261) & -0.00003(0) & 0.00003(274) &-0.00005(0) & 0.0000(274) & -0.00006(0) \\
			\hline
		\end{tabular}
	\end{center}
%	\caption{\label{tab:pressummary2} Summary of the estimated shear pressure (negative shear stress) in the fluid region, as the average over the planes $2-6$. The system at $P\simeq7.7$ is the case at constant volume. The $n$ values in brackets are twice the standard error and is the error in the last $n$ digits of the shear pressures quoted.}
\end{table}
% fig 9

% CONCLUSIONS
\section{Conclusions}

We have derived the analytical TTCF expression for boundary sheared systems at constant pressure, and have shown that it is equal to that at constant volume as long as the barostat is modelled via a conservative force. We have studied systems at different pressure and for various driving boundary velocities, at realistic strain rates. For all the quantities monitored, we observed an increasing level of noise in the DAV signal, making the signal-to-noise ratio rapidly decay to zero. On the other hand, the signal-to-noise ratio remained constant with the TTCF method and was substantially higher, indicating that accurate measurements can be obtained for much weaker external driving forces. Our results show that the correlation with the dissipation function and both the shear pressure (negative of the shear stress) and the streaming velocity does not substantially decay when moving far from the wall. Although the limited size of the systems along the $x$-direction could affect the final value of the shear pressure, we have shown that the TTCF method can be applied in the study of friction in nanochannels at realistic rates of strain. The generality of the TTCF approach, and its simplicity, make it readily expandable to more complex systems such as molecular fluids and mixtures and opens up a new frontier in the use of NEMD simulation to study the underlying physical processes that determine the tribology of highly confined systems at realisable strain rates.

\section{Acknowledgements}
The authors would like to thank Prof. D. M. Heyes (Department of Mechanical Engineering, Imperial College London, UK) for useful discussions. We also thank the Australian Research Council for a grant obtained through the Discovery Projects Scheme (DP200100422) and the Royal Society for support via International Exchanges grant $IES\ R3\ 170233$. J.P.E. and D.D. acknowledge the financial support of the Engineering and Physical Sciences Research Council (EPSRC) via EP/ N025954/1 and EP/P030211/1. J.P.E. was supported by the Royal Academy of Engineering through the Research Fellowships scheme. We acknowledge the Swinburne OzSTAR Supercomputing facility and Imperial College London Research Computing Service for providing computational resources for this work. 

% Data availability
\section{Data availability statement}
The data that support the findings of this study are available from the corresponding author or by emailing tribology@imperial.ac.uk upon reasonable request.

% Appendix
\section{Appendix}

In this Appendix, we provide a derivation of the dissipation function for barostatted systems, and we show that it is equivalent to that of a system with fixed walls. 
We recall the equations of motion
\begin{equation}
\begin{split}
\dot{\textbf{r}}^f_i&=\dfrac{\textbf{p}_i^{f}}{m_i}\\
\dot{\textbf{p}}^f_i&=\textbf{F}_i^{2B}
\end{split}
\end{equation}

\begin{equation}
\begin{split}
\dot{\textbf{r}}^w_i&=\dfrac{\textbf{p}_i^{w}}{m_i}\\
\dot{\textbf{p}}^w_i&=\textbf{F}_i^{2B}+\textbf{F}_i^{H}-\alpha\textbf{p}^w_i
\end{split}
\end{equation}
\begin{equation}
\begin{split}
\dot{\textbf{r}}^l_i&=\biggl(v\;\;,\;\;\dfrac{p_y^l}{m_i}\;\;,\;\;0\biggr)\\
\dot{\textbf{p}}^l_i&=\biggl(0\;\;,\;\;-F^{ext}-\dfrac{\sum_i^{N^l}F_{yi}^H}{N^l}\;\;,\;\;0\biggr),
\end{split}
\end{equation}
where, $\textbf{F}_i^H=-k(\textbf{r}_i^w-\textbf{r}_i^l)$ and the sign of the external force is negative because we are moving the upper wall (i.e. the sign depends on the direction of the wall that is being driven). The difference between upper and lower wall velocity is ignored, since it does not affect the calculations.  
\\
The total energy $U=K+V$ can now be expressed as
\begin{equation}
\begin{split}
K&=\sum_i^{N^f}\dfrac{\textbf{p}_i^{2f}}{2m_i}+\sum_i^{N^w}\dfrac{\textbf{p}_i^{2w}}{2m_i} +N^l\dfrac{p_y^{2l}}{2m_i}\\
V&=\sum_i^{N^t}\sum^{N^t}_{j>i}\phi_{ij}^{2B}+\sum_i^{N^w}\phi_i^{H}+F^{ext}\sum_i^{N^l}(r_{yi}^l-r_{yi0}^l)
\end{split}
\end{equation}
with $N^t=N^f+N^w$, $\phi_i^{H}=\frac{1}{2}k(\textbf{r}_i^w-\textbf{r}_i^l)^2$ and $r_{yi0}^l$ the arbitrary reference position of the lattice site $i$.
The total time derivative of the energy $U$, which is proportional to the dissipation function we require ($\Omega$), is
\begin{equation}
\begin{split}
\dfrac{\text{d}U}{\text{d}t}=\dfrac{\partial U}{\partial t}+&\sum_i^{Nf}\biggl(\dfrac{\partial}{\partial\textbf{r}^f_i}\cdot\dot{\textbf{r}}^f_i+\dfrac{\partial}{\partial\textbf{p}^f_i}\cdot\dot{\textbf{p}}^f_i\biggr)U+\\
&\sum_i^{Nw}\biggl(\dfrac{\partial}{\partial\textbf{r}^w_i}\cdot\dot{\textbf{r}}^w_i+\dfrac{\partial}{\partial\textbf{p}^w_i}\cdot\dot{\textbf{p}}^w_i\biggr)U+\sum_i^{Nl}\biggl(\dfrac{\partial}{\partial\textbf{r}^l_i}\cdot\dot{\textbf{r}}^l_i+\dfrac{\partial}{\partial\textbf{p}^l_i}\cdot\dot{\textbf{p}}^l_i\biggr)U.
\end{split}
\end{equation}
Recalling that for any generic particle $k$ we have
\begin{equation}
\begin{split}
\dfrac{\partial}{\partial\textbf{r}_k}\biggl(\sum_i^{Nt}\sum^{Nt}_{j>i}\phi_{ij}^{2B}\biggr)&=-\textbf{F}_k^{2B} \\
\dfrac{\partial}{\partial\textbf{r}^w_k}\biggl(\sum_i^{Nw}\phi_i^{H}\biggr)&=-\textbf{F}_k^{H} \\
\dfrac{\partial}{\partial\textbf{r}^l_l}\biggl(\sum_i^{Nw}\phi_i^{H}\biggr)&=\textbf{F}_k^{H},
\end{split}
\end{equation}
then the total time derivatives of the kinetic and potential energy are
\begin{equation}
\begin{split}
\dfrac{\text{d}K}{\text{d}t}&=\sum_i\biggl(\dfrac{\partial}{\partial\textbf{p}_i}\cdot\dot{\textbf{p}}_i\biggr)K= \sum_i^{Nf}\textbf{p}^f_i\cdot(\textbf{F}_i^{2B})+\sum_i^{Nw}\textbf{p}^w_i\cdot(\textbf{F}_i^{2B}+\textbf{F}_i^{H})+\sum_i^{N^l}\dfrac{p^l_y}{m_i}(-F^{ext}-\dfrac{\sum_iF_{yi}^H}{N^l})
\end{split}
\end{equation}
and
\begin{equation}
\begin{split}
\dfrac{\text{d}V}{\text{d}t}&=\sum_i\biggl(\dfrac{\partial}{\partial\textbf{r}_i}\cdot\dot{\textbf{r}}_i\biggr)V= \sum_i^{Nf}\dfrac{\textbf{p}^f_i}{m_i}\cdot(-\textbf{F}_i^{2B})+\sum_i^{Nw}\dfrac{\textbf{p}^w_i}{m_i}\cdot(-\textbf{F}_i^{2B}-\textbf{F}_i^{H})+\sum_i^{Nl}\dot{\textbf{r}}_i^l\cdot(\textbf{F}_i^H)+F^{ext}\sum_i^{N^l}\dfrac{p^l_{y}}{m_i}.
\end{split}
\label{eqn:kinderiv}
\end{equation}
With the assumption that all the masses $m_i$ are identical and equal to $m$, Eq. \eqref{eqn:kinderiv} can be simplified and combined with the kinetic contribution into
\begin{equation}
\dfrac{\text{d}U}{\text{d}t}=\dfrac{N^l p_y^l}{m}\biggl(-\sum_i^{N^l}\dfrac{F_{yi}^H}{N^l}\biggr)+\sum_i^{N^l}\dot{\textbf{r}}_i^l\cdot\textbf{F}_i^H.
\end{equation}
Recalling that for a particle $k$ of the lattice 
\begin{equation}
\dot{\textbf{r}}_k^l\cdot\textbf{F}_k^H=\biggl(v\;\;,\;\;\dfrac{p_y^l}{M}\;\;,\;\;0\biggr)\cdot\biggl(F^{H}_{xi}\;\;,\;\;F^{H}_{yi}\;\;,\;\;F^{H}_{zi}\biggr)^{T}=F^{H}_{xi}v+F^{H}_{yi}\dfrac{p_y^l}{m}
\end{equation}
we finally have
\begin{equation}
\dfrac{\text{d}U}{\text{d}t} = -\dfrac{p_y^l}{m}\sum_i^{N^l} F^{H}_{yi}+\dfrac{p_y^l}{m}\sum_i^{N^l}F^{H}_{yi}+v\sum_i^{N^l}F^{H}_{xi}=-\sum_i^{Nl}k(r_{xi}^w-r_{xi}^l)v,
\end{equation}
hence 
\begin{equation}
\Omega = \beta \frac{dU}{dt} = \beta \sum_i^{Nl}k(r_{xi}^w-r_{xi}^l)v.
\end{equation}
We note that the dissipation function derived is identical to that of a system with fixed walls. This property follows immediately from the conservative nature of the forces involved in barostatting the system, and hence its generality is quite wide and immediately applicable to several other methods of pressure control for inhomogeneous systems.

% REFERENCES
\bibliography{bibliography}
\bibliographystyle{ieeetr}

\end{document}